\documentclass[12pt]{iopart}
\usepackage{graphicx}

\begin{document}
\def \ee {\varepsilon}
\thispagestyle{empty}
\title[]{Problems in the theory of thermal Casimir force between
dielectrics and semiconductors}

\author{
G~L~Klimchitskaya${}^{1,2}$ and
B~Geyer${}^{1}$
}

\address{${}^1$Center of Theoretical Studies and Institute for Theoretical
Physics, Leipzig University,
D-04009, Leipzig, Germany}

\address{$^2$North-West Technical University, Millionnaya St. 5,
St.Petersburg, 191065, Russia
}

\begin{abstract}
The application of the Lifshitz theory to describe the thermal Casimir
force between dielectrics and semiconductors is considered.
It is shown that for all true dielectrics (i.e., for all materials
having zero conductivity at zero temperature) the inclusion of a
nonzero conductivity arising at nonzero temperature into the model
of dielectric response leads to the violation of the Nernst heat theorem.
This result refers equally to simple insulators, intrinsic semiconductors,
Mott-Hubbard dielectrics and doped semiconductors with doping
concentration below a critical value. We demonstrate that in the
insulator-metal transition the Casimir free energy changes abruptly
irrespective of whether the conductivity changes continuously or
discontinuously. The application of the Lifshitz formula to polar
dielectrics results in large thermal correction that is linear in
temperature. {A rule is formulated on how to apply the Lifshitz theory to real
materials in agreement with thermodynamics and
experiment.}
\end{abstract}

\section{Introduction}

The last ten years are marked by quick progress in experimental
investigation of the Casimir effect \cite{1}. The early stages of this
process (reflected in review \cite{2}) were followed by the measurement
of the Casimir force
in the original configuration of two parallel plates  \cite{3}
and first experimental demonstration of the lateral Casimir force.
Thereafter the precise measurements by using the micromechanical torsional
oscillator \cite{5,6,7} and first experiments on measuring the Casimir
force between metal and semiconductor test bodies \cite{8,9} were
performed. It was demonstrated also that the force between metal and
semiconductor can be controlled by the illumination of a semiconductor
plate with laser pulses \cite{10}.

Intensive experimental work created demands to theoretical computations
of the Casimir force between real material bodies used in the
laboratory setups.
The Casimir effect arises due to quantum fluctuations of the
electromagnetic field \cite{10a}.
{ The recognized basic theory} of both the van der Waals
and Casimir force is the Lifshitz theory \cite{11,12}.
In the framework of this theory, material properties are described by the
dielectric permittivity $\varepsilon(\omega,T)$ depending on the
frequency $\omega$ and the temperature $T$. The free energy per unit area
of the van der
Waals and Casimir interaction between two thick plane parallel
plates at a separation $a$ in thermal equilibrium is given by \cite{2,11,12}
\begin{eqnarray}
&&
{\cal F}(a,T)=\frac{k_BT}{2\pi}\sum\limits_{l=0}^{\infty}
\left(1-\frac{1}{2}\delta_{0l}\right)
\int_{0}^{\infty}k_{\bot}\,dk_{\bot}
\nonumber \\
&&\phantom{aaa}
\times\left\{\ln\left[1-r_{\rm TM}^2({\rm i}\xi_l,k_{\bot})
e^{-2aq_l}\right]+
\ln\left[1-r_{\rm TE}^2({\rm i}\xi_l,k_{\bot})
e^{-2aq_l}\right]\right\}.
\label{eq1}
\end{eqnarray}
\noindent
Here
$k_B$ is the Boltzmann constant, {$\xi_l=2\pi k_B Tl/\hbar$}
with {$l=0,\,1,\,2,\,\ldots$}
are the Ma\-tsu\-ba\-ra frequencies and
the reflection coefficients for the two independent
polarizations of the electromagnetic field (transverse
magnetic and transverse electric) are given by
\begin{eqnarray}
&&
r_{\rm TM}({\rm i}\xi_l,k_{\bot})=
\frac{\varepsilon_lq_l-k_l}{\varepsilon_lq_l+k_l},
\qquad
r_{\rm TE}({\rm i}\xi_l,k_{\bot})=
\frac{k_l-q_l}{k_l+q_l},
\label{eq2} \\
&&q_l=\sqrt{k_{\bot}^2+\frac{\xi_l^2}{c^2}}, \quad
k_l=\sqrt{k_{\bot}^2+\varepsilon_l\frac{\xi_l^2}{c^2}},
\quad \varepsilon_l=\varepsilon({\rm i}\xi_l,T)
\nonumber
\end{eqnarray}
\noindent
($k_{\bot}=|\mbox{\boldmath$k$}_{\bot}|$ is the projection
of a wave vector on the plane of the plates).
It should be noted that all Matsubara frequencies with $l\geq 1$ are
rather high at all accessible temperatures. As an example, at room
temperature $T=300\,$K it holds
\begin{equation}
\xi_1=2.47\times 10^{14}\,\mbox{rad/s}, \qquad \xi_l=l\xi_1.
\label{eq3}
\end{equation}

However, the application of the Lifshitz theory for the calculation
of the thermal Casimir force between real metals leads to serious
problems. It was shown { that the usual description} of a metal by means
of the Drude dielectric function results in the violation of the
third principle of thermodynamics (the Nernst heat theorem)
in the case of perfect crystal lattice \cite{13} and in
contradictions with experiment \cite{5,6,7} (see review in \cite{14}).
The application of the Lifshitz theory at nonzero temperature to
dielectric materials taking into account their conductivity at
zero frequency also leads to the violation of Nernst's theorem
\cite{15,16}. Leaving aside the case of metals (see \cite{17,18}
for further discussion), we concentrate in this paper on the problems
of the Lifshitz theory arising when it is applied to real
dielectrics and semiconductors.

Below we demonstrate that for all true dielectrics (i.e., for
materials having zero conductivity at $T=0$) the account of nonzero
conductivity arising at $T>0$ leads to a violation of the Nernst
heat theorem in the Lifshitz theory.
In particular we show that for doped semiconductors with
sufficiently low doping concentration (i.e., lower than the
critical concentration above which the conductivity is of
metallic type) the account of conductivity at zero frequency
violates the Nernst heat theorem as well. For doped Si samples
with low doping concentration the inclusion of conductivity
arising at $T>0$ has been rejected experimentally at 95\%
confidence level \cite{10}. According to the obtained results,
in the insulator-metal transition the Casimir free energy and
force change abruptly irrespective of whether the conductivity
changes continuously or discontinuously. The account of orientation
polarization in polar dielectrics results in a large thermal
{ correction, being linear in the temperature, to the Casimir force
at separations of the order of hundreds of nanometers}. Arguments are
presented that this effect is nonphysical.

The paper is organized as follows. In Section 2 we discuss the main
characteristic features of simple dielectrics, metals, semimetals
and doped semiconductors. Section 3 is devoted to the violation of the
Nernst theorem for all materials whose conductivity at zero temperature
is equal to zero if at $T\neq 0$ their conductivity is included into
the model of
dielectric response in the Lifshitz theory. The Casimir effect in
insulator-metal transition is considered in Section 4. In Section 5
we discuss problems which arise for polar dielectrics.
Section 6 contains our conclusions and discussion.

\section{Free charge carriers in different materials}

It is common knowledge that at nonzero temperature all materials
contain some amount of free charge carriers. For materials with
very low charge carrier density (insulators) the dielectric
permittivity is usually represented in the form
\begin{equation}
\varepsilon(\omega)=1+
\sum_j\frac{g_j}{\omega_j^2-\omega^2-{\rm i}\gamma_j\omega},
\label{eq4}
\end{equation}
\noindent
where $\omega_j\neq 0$ are the oscillator frequencies,
$g_j$ are the oscillator strengths and $\gamma_j$ are the damping
parameters (note that in general case  the parameters of oscillators
may depend on temperature but this minor dependence can be
neglected).
From (\ref{eq4}) the dielectric
permittivity at zero frequency is given by
\begin{equation}
\varepsilon_0\equiv\varepsilon(0)=1+
\sum_j\frac{g_j}{\omega_j^2}<\infty.
\label{eq5}
\end{equation}

Equation (\ref{eq4}) does not take free charge carriers into account.
{This means  that} in this simple model the conductivity
of insulator is assumed to be equal to zero at any temperature.
However, at {$T\neq 0$} all insulators possess some nonzero
con\-duc\-ti\-vi\-ty, {$\sigma_0\neq 0$}, and respective finite resistivity
$\rho=1/\sigma_0$. As an example, at room temperature ($T=300\,$K)
resistivity of different insulators can vary over a wide range
from about $10^8$ to about $10^{17}\,\Omega\,$cm. Resistivity of metals
varies in the range from $10^{-6}$ to about $10^{-4}\,\Omega\,$cm.
By convention the range of resistivities from about $10^{-3}$ to
$10^{7}\,\Omega\,$cm is attributed to semiconductors.
The dielectric permittivity of insulators and semiconductors at
nonzero temperature can be represented in the form \cite{20}
\begin{equation}
\tilde\varepsilon(\omega,T)=\varepsilon(\omega)+
{\rm i}\frac{4\pi\sigma_0(T)}{\omega},
\label{eq6}
\end{equation}
\noindent
where $\varepsilon(\omega)$ is given in (\ref{eq4}).

The characteristic properties of different materials are determined
by the behavior of the density $N$  of one-electron states as a function of
energy $E_e$ at zero temperature \cite{19}.
As was stated in \cite{19}, the most fundamental property of true
dielectrics separating them from metals is that the former possess
zero conductivity, $\sigma_0=0$, at zero temperature.
In figure 1(a) we present the typical functional form of $N(E_e)$
for insulators and intrinsic (i.e., undoped) semiconductors \cite{21}.
In this figure, all states in the shaded region are filled (this is
the valence band) and all states in the nonshaded region are empty
(the conduction band). The Fermi energy, $E_F$, separates
filled and empty states. As is seen in figure 1(a), here
$N(E_F)=0$. From this it follows that $\sigma_0(T=0)=0$ for both
insulators and intrinsic semiconductors. The universal behavior
of their conductivity at $T\neq 0$ is given by
\begin{equation}
\sigma_0(T)\sim\exp\left(-\frac{\Delta}{2k_BT}\right),
\label{eq7}
\end{equation}
\noindent
where $\Delta$ is the bandgap. By convention the material is
called insulator if $\Delta\geq 2-3\,$eV and intrinsic semiconductor
if $\Delta< 2-3\,$eV \cite{21}. For comparison in figure 2(b) we
show schematically the typical functional form of $N(E_e)$ for
metals. Here, $N(E_F)\neq 0$ and the conductivity at zero temperature
is not equal to zero, $\sigma_0(T=0)\neq 0$.
\begin{figure*}[t]
\vspace*{-14.4cm}
\hspace*{-0.5cm}\includegraphics{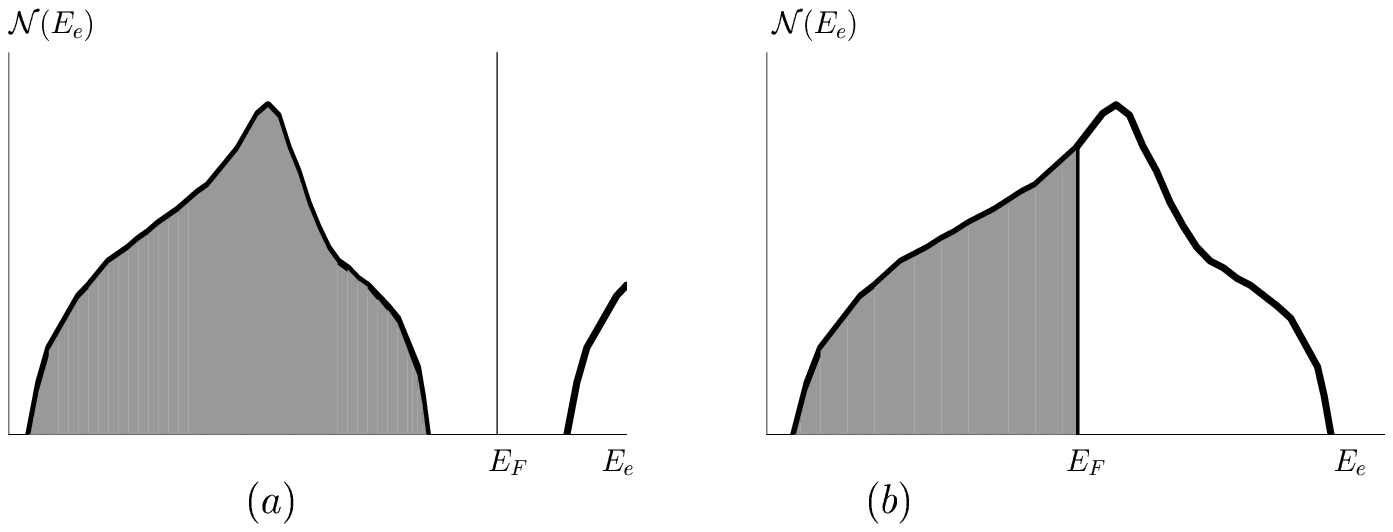}
\vspace*{-11.cm}
\caption{
Density of states $N$ (a) for insulators and intrinsic semiconductors
and (b) for metals as a function of energy $E_e$.
The filled states are shaded, $E_F$ is the Fermi energy.
}
\end{figure*}

There are, however, materials which are characterized by a nonzero
density of states at $E=E_F$, but have zero conductivity at
zero temperature. To illustrate this we consider the typical
behavior of the density of states at $T=0$, as shown in figure 2(a).
Here, the valence and conduction bands overlap and the resulting band
is restricted by the solid line. The band structure of figure 2(a)
in fact describes  two different types of materials: semimetals and
Mott-Hubbard dielectrics \cite{21}. For semimetals it holds
$N(E_F)\neq 0$ and $\sigma_0(T=0)\neq 0$ like for usual metals \cite{22}.
As to Mott-Hubbard dielectrics, they are characterized by
$N(E_F)\neq 0$ but $\sigma_0(T=0)=0$. This is explained by the fact
that for such materials {the one-electron
approximation works rather bad}
 and electron correlations play an important role \cite{21}.
At $T>0$ Mott-Hubbard dielectrics have some nonzero conductivity
that depends on the temperature as
\begin{equation}
\sigma_0(T)\sim\exp\left(-\frac{C}{k_BT}\right),
\label{eq8}
\end{equation}
\noindent
where the parameter $C$ {has a different physical meaning} than  $\Delta$
in (\ref{eq7}).

One more type of materials are doped semiconductors. They are obtained
from intrinsic semiconductors by the inclusion of some foreign atoms
in their crystal lattice. The typical density of states for these
materials is shown in figure 2(b) ($n$-type semiconductor), where
the first zone containing the Fermi energy is the impurity band.
The second (empty) zone is the conduction band of the
intrinsic semiconductor. The valence band of the intrinsic semiconductor
is not shown [it is the same as in figure 1(a)]. As is seen in
figure 2(b), $N(E_F)\neq 0$. It is important that for doped
semiconductors $\sigma_0(T=0)\neq 0$ for doping concentration
$n>n_{cr}$, where $n_{cr}$ is the so-called {\it critical} doping
concentration, and  $\sigma_0(T=0)=0$ for $n<n_{cr}$.
This can be explained as follows \cite{23}. In the perfect crystal
lattice of an intrinsic semiconductor delocalization of electrons is
caused by the periodicity of the lattice. Impurity centers are
distributed randomly. The one-electron states which form the
impurity zone are of different nature depending on whether
$n<n_{cr}$ or $n>n_{cr}$. If $n<n_{cr}$, electrons are localized
in the vicinity of impurity centers. For the localized electron
states, conductivity at $T=0$ is equal to zero (in the same way
as for the delocalized electrons states due to the perfect lattice
of an intrinsic semiconductor). However, when $n>n_{cr}$ the
electron states of impurities overlap and due to this become
delocalized. Thus, they are of the same kind as electron states in
metals leading to $\sigma_0(T=0)\neq 0$. An example of this
situation is given by Si doped with P. For this case
$n_{cr}\approx 3.7\times 10^{18}\,\mbox{cm}^{-3}$. If $n<n_{cr}$
the conductivity of P-doped Si at $T=0$ is equal to zero.
At sufficiently low $T$ it is given by (\ref{eq8}) with some
constant $C$ \cite{23}.
\begin{figure*}[t]
\vspace*{-14.4cm}
\hspace*{-1cm}\includegraphics{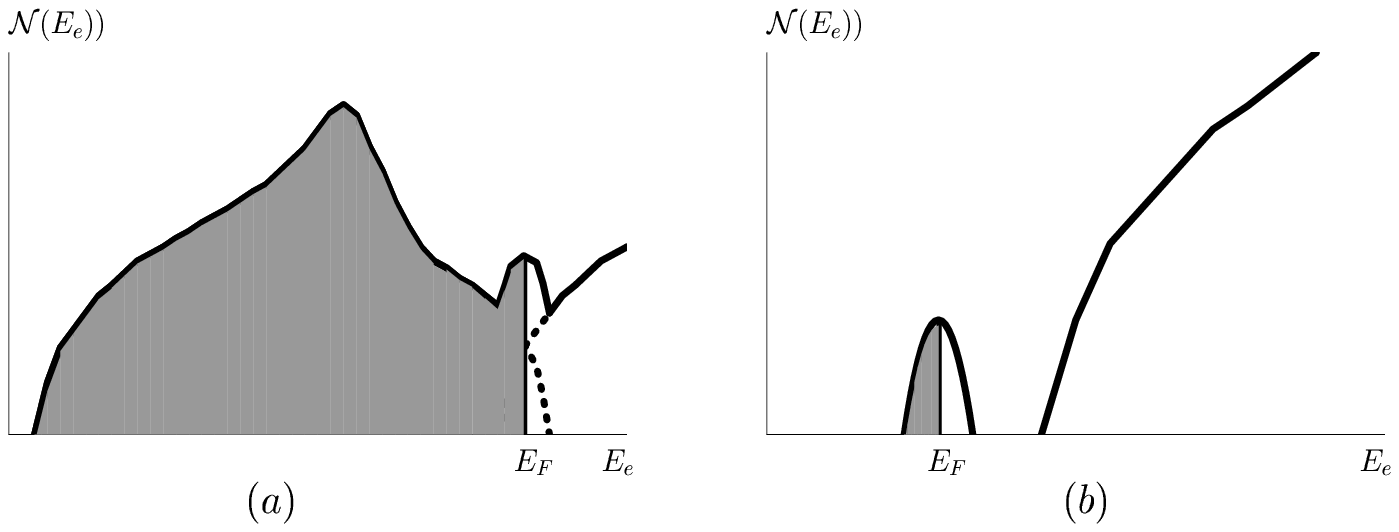}
\vspace*{-11.cm}
\caption{
Density of states $N$ (a) for semimetals and Mott-Hubbard dielectrics
and (b) for $n$-type semiconductors where only the impurity band
is shown. The filled states are shaded, $E_F$ is the Fermi energy.
}
\end{figure*}

We emphasize that for all materials whose conductivity at $T=0$
is equal to zero (insulators, intrinsic semiconductors,
Mott-Hubbard dielectrics, doped semiconductors with $n<n_{cr}$)
the low-temperature behavior of their conductivity obeys equations
(\ref{eq7}), (\ref{eq8}), i.e., $\sigma_0$ vanishes exponentially
fast with the temperature.

\section{Thermodynamic test for the Lifshitz theory of the Casimir
force between dielectrics and semiconductors}

Here we consider the low-temperature asymptotic
behavior of the Lifshitz
formula (\ref{eq1}), (\ref{eq2}) for the Casimir free energy in
combination with the dielectric permittivities (\ref{eq4}) and
 (\ref{eq6}). We also calculate the respective Casimir entropy and
check whether or not the Nernst heat theorem is satisfied.

It is convenient to introduce the dimensionless variables
\begin{equation}
y=2q_la, \qquad \zeta_l=\frac{\xi_l}{\omega_c}=\tau l, \qquad
\omega_c=\frac{c}{2a}, \qquad
\tau=\frac{4\pi k_BaT}{\hbar c}.
\label{eq9}
\end{equation}
\noindent
By using the Abel-Plana formula \cite{2}
\begin{equation}
\sum_{l=0}^{\infty}\left(1-\frac{1}{2}\delta_{l0}\right)F(l)=
\int_{0}^{\infty}F(t)dt+{\rm i}\int_{0}^{\infty}dt
\frac{F({\rm i}t)-F(-{\rm i}t)}{{\rm e}^{2\pi t}-1},
\label{eq10}
\end{equation}
\noindent
the Casimir free energy can be represented as the sum of the
energy at $T=0$ and the thermal correction to it
\begin{equation}
{\cal F}(a,T)=E(a)+\Delta_T{\cal F}(a,T),
\label{eq11}
\end{equation}
\noindent
where
\begin{eqnarray}
&&
E(a)=\frac{\hbar c}{32\pi^2 a^3}\int_{0}^{\infty}d\zeta
\int_{\zeta}^{\infty}dyf(\zeta,y),
\nonumber \\
&&
f(\zeta,y)=
y\left\{\ln\left[1-r_{\rm TM}^2({\rm i}\zeta,y){\rm e}^{-y}\right]
+\ln\left[1-r_{\rm TE}^2({\rm i}\zeta,y){\rm e}^{-y}\right]\right\},
\nonumber \\
&&
\Delta_T{\cal F}(a,T)=\frac{{\rm i}\hbar c\tau}{32\pi^2a^3}
\int_{0}^{\infty}dt
\frac{F({\rm i}t\tau)-F(-{\rm i}t\tau)}{{\rm e}^{2\pi t\tau}-1},
\nonumber \\
&&
F(x)\equiv\int_{x}^{\infty}dyf(x,y).
\label{eq12}
\end{eqnarray}
\noindent
The reflection coefficients expressed in terms of dimensionless
variables are given by
\begin{equation}
\hspace*{-1.7cm}
r_{\rm TM}({\rm i}\zeta,y)=\frac{\varepsilon y-
\sqrt{y^2+\zeta^2(\varepsilon-1)}}{\varepsilon y+
\sqrt{y^2+\zeta^2(\varepsilon-1)}},
\qquad
r_{\rm TE}({\rm i}\zeta,y)=\frac{\sqrt{y^2+\zeta^2(\varepsilon-1)}
-y}{\sqrt{y^2+\zeta^2(\varepsilon-1)}+y}.
\label{eq12a}
\end{equation}
\noindent
Now we substitute $\varepsilon$ from (\ref{eq4}) in the reflection
coefficients (\ref{eq12a}).
To obtain the asymptotic behavior of the thermal correction,
$\Delta{\cal F}(a,T)$, at $\tau\ll 1$, we expand the function
$f(x,y)$ in (\ref{eq12}) in powers of $x=t\tau$. The subsequent
integration of this expansion with respect to $y$ from $x$ to infinity
results in
\begin{eqnarray}
&&
F({\rm i}x)-F(-{\rm i}x)=\frac{8{\rm i}bx}{\varepsilon_0^2-1}
{\rm Li}_2(r_0^2)+\frac{{\rm i}\pi}{2}r_0^2(\varepsilon_0+1)x^2
\label{eq13} \\
&&\phantom{aaaaaaaaaaaaaaaaaaaaaaaa}
-240{\rm i}C_4x^3+O(x^4),
\nonumber
\end{eqnarray}
\noindent
where ${\rm Li}_n(z)$ is the polylogarithm function,
$C_4$ is some coefficient and the following
notations are used
\begin{equation}
b=b(a)=\sum_{j}\frac{g_j\gamma_j\omega_c}{\omega_j^4}, \qquad
r_0=\frac{\varepsilon_0-1}{\varepsilon_0+1}.
\label{eq14}
\end{equation}
\noindent
Substituting (\ref{eq13}) in (\ref{eq12}) and performing integration
with respect to $t$ from zero to infinity, we obtain
\begin{eqnarray}
&&
{\cal F}(a,T)=E(a)-\frac{\hbar c}{32\pi^2a^3}\left[
\frac{b{\rm Li}_2(r_0^2)}{3(\varepsilon_0^2-1)}\tau^2+
\frac{\zeta(3)r_0^2(\varepsilon_0+1)}{8\pi^2}\tau^3\right.
\nonumber \\
&&\phantom{aaaaaaaaaaaaaaaaaa}
\left. -C_4\tau^4+O(\tau^5)
\vphantom{\frac{b{\rm LI}_2(r_0^2)}{3(\varepsilon_0+1)}}
\right],
\label{eq15}
\end{eqnarray}
\noindent
where $\zeta(z)$ is the Riemann zeta function.
The explicit expression for the coefficient $C_4$ can be found as
in \cite{15} by considering the Lifshitz formula for the Casimir
pressure. In the case $\gamma_j=0$ the result is
\begin{equation}
C_4=\frac{1}{720}(\varepsilon_0^{1/2}-1)(\varepsilon_0^2+
\varepsilon_0^{3/2}-2).
\label{eq15a}
\end{equation}
\noindent
 From (\ref{eq15})
the Casimir entropy is given by the expression
\begin{equation}
\hspace*{-2.2cm}
S(a,T)=-\frac{\partial{\cal F}(a,T)}{\partial T}=
\frac{k_B\tau}{8\pi a^2}\left[
\frac{2b{\rm Li}_2(r_0^2)}{3(\varepsilon_0^2-1)}
+
\frac{3\zeta(3)r_0^2(\varepsilon_0+1)}{8\pi^2}\tau
 -4C_4\tau^2+O(\tau^3)
\right].
\label{eq16}
\end{equation}
\noindent
As is seen from (\ref{eq16}), $S(a,T)$ goes to zero when $T$ goes
to zero, i.e., the Nernst heat theorem is satisfied when the
dielectric permittivity is given by (\ref{eq4}) with a finite static
value (\ref{eq5}).

Now we take into account the conductivity of true dielectrics that
arises at $T>0$. In this case we should replace the dielectric
permittivity (\ref{eq4}) with the dielectric permittivity (\ref{eq6}),
where at low temperatures the conductivity $\sigma_0(T)$ decreases
exponentially with $T$ as given in (\ref{eq7}), (\ref{eq8}).
We remind that this universal behavior is relevant to all materials
with $\sigma_0(T=0)=0$, i.e., for insulators, intrinsic semiconductors,
Mott-Hubbard dielectrics and doped semiconductors with $n<n_{cr}$.
{}From (\ref{eq6}) it follows
\begin{equation}
\tilde\varepsilon_l=\tilde\varepsilon({\rm i}\xi_l,T)=
\varepsilon_l+\frac{\beta(T)}{l},
\label{eq17}
\end{equation}
\noindent
where $\beta(T)=2\hbar\sigma_0(T)/(k_BT)$.
{}From (\ref{eq7}), (\ref{eq8}) we conclude that at sufficiently low $T$
it holds $\beta(T)\ll 1$ for all materials with $\sigma_0(T=0)=0$.
Repeating the above calculation of the Casimir free energy at low
temperatures using the Lifshitz formula, one arrives at the result
(see \cite{15} for details)
\begin{equation}
\tilde{\cal F}(a,T)={\cal F}(a,T)-\frac{k_BT}{16\pi a^2}
\left[\zeta(3)-{\rm Li}_3(r_0^2)+R(\tau)\right],
\label{eq18}
\end{equation}
\noindent
where ${\cal F}(a,T)$ is given in (\ref{eq15}) and $R(\tau)$
decreases exponentially when $T$ vanishes. From (\ref{eq18})
the Casimir entropy at $T=0$ is given by
\begin{equation}
\tilde{S}(a,0)=\frac{k_B}{16\pi a^2}
\left[\zeta(3)-{\rm Li}_3(r_0^2)\right]>0
\label{eq19}
\end{equation}
\noindent
in violation of the Nernst theorem.

Thus, inclusion of nonzero conductivity arising at $T>0$ into the model
of dielectric response for all materials possessing zero conductivity
at zero temperature leads to contradictions between the Lifshitz theory and
thermodynamics. That is why for true dielectrics conductivity must be
disregarded in theoretical computations. This conclusion has been
already confirmed experimentally in the measurements of the difference
Casimir force between an Au sphere and a Si plate illuminated with laser pulses
\cite{10}. In figure 3(a) we plot the difference of the Casimir force
between a sphere and a plate, $\Delta F$, when the laser light is on and off,
as a function of separation. Mean experimentally measured difference data
are shown as dots. Solid line shows the theoretical results computed using
the dielectric permittivity (\ref{eq4}) in the absence of laser pulse
(in this case $n<n_{cr}$). Dashed line was obtained using the dielectric
permittivity (\ref{eq6}) in the absence of laser light, i.e., taking
conductivity into account. As is seen in figure 3(a), the solid line
is in good agreement with data, whereas the dashed line is experimentally
excluded. For illustrative purposes, the same data and theories are
presented over a more narrow separation interval in figure 3(b) with
indication of experimental errors found at  95\% confidence level.
The solid and dashed lines have the same meaning as in figure 3(a)
representing the theoretical force differences computed using the
dielectric permittivities (\ref{eq4}) and (\ref{eq6}), respectively,
in the absence of laser light. It is clearly seen { that a theory
taking
into account} the conductivity of doped Si with doping concentration
below its critical value is experimentally excluded. {But, a theory which}
disregards this conductivity of Si is in good agreement with data.
\begin{figure*}[t]
\vspace*{-11.cm}
\hspace*{-2.5cm}\includegraphics{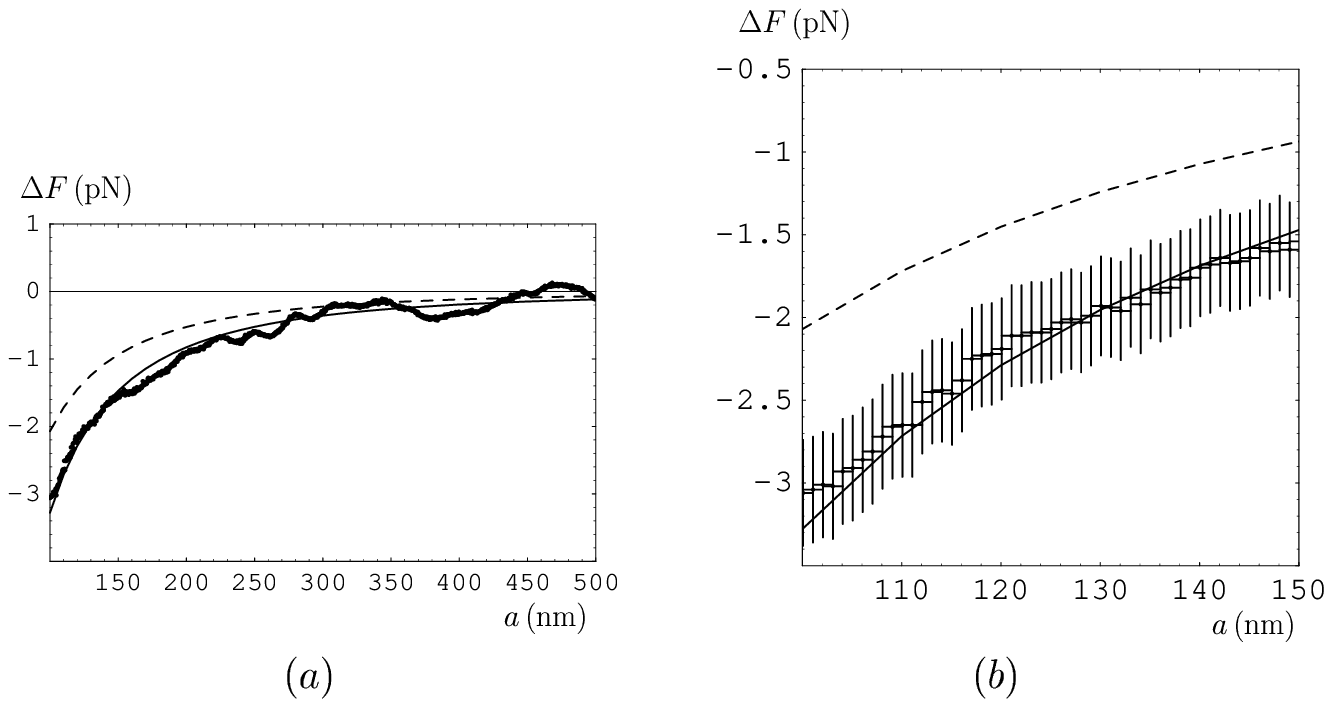}
\vspace*{-12.cm}
\caption{
Differences of the Casimir forces between Au sphere and Si plate illuminated
with laser pulses with light on and off versus separation.
Solid and dashed lines represent theoretical differences computed
at $T=300\,$K disregarding and taking into account the conductivity
of Si plate in the absence of light. Mean experimental differences
are shown as dots. (a) Complete data over the entire measurement range.
(b) Data over a more narrow separation interval with experimental
errors determined at  95\% confidence level shown as crosses.
}
\end{figure*}

\section{The Casimir effect in the insulator-metal transition}

Insulator-metal transition is the phase transition leading to a change
of the character (and magnitude) of conductivity with the change of
temperature, pressure, magnetic field or composition. A familiar example
is an abrupt transition from the monoclinic phase at room temperature
to tetragonal phase at $T>340\,$K in VO${}_2$ \cite{24}. Such a
transition leads to a change of conductivity of order $10^4$.
Recently an experiment has been proposed \cite{25} { measuring
 the change} of the Casimir force acting between an Au coated sphere
and a  VO${}_2$ film deposited on sapphire substrate which undergoes the
insulator-metal transition with the increase of temperature.
Before the phase transition, in accordance with the results of Section 3,
the conductivity properties of  VO${}_2$ should not be included in the
model of dielectric response. Thus, the results of this experiment
could be used as one more fundamental test of the Lifshitz theory in
application to real materials.

One more example is the insulator-metal phase transition which takes place in
$n$-Si doped by P with the increase of doping concentration.
In this case the critical concentration is
$n_{cr}\approx 3.7\times 10^{18}\,\mbox{cm}^{-3}$ \cite{23}. We consider the
doping concentrations
$n_1\approx 2.5\times 10^{18}\,\mbox{cm}^{-3}$ (the respective
resistivity is $\rho_1\approx 2\times 10^{-2}\,\Omega\,$cm \cite{26})
just before the phase transition and
$n_2\approx 5\times 10^{18}\,\mbox{cm}^{-3}$
($\rho_2\approx 1\times 10^{-2}\,\Omega\,$cm) after the phase transition.
It is easily seen that this transition has a pronounced effect only
on the zero contribution to the Lifshitz formula (\ref{eq1}) and
practically does not influence the contributions of all Matsubara
frequencies with $l\geq 1$. It is interesting to find the difference
between the Casimir free energies after, ${\cal F}_2(a,T)$, and
before, ${\cal F}_1(a,T)$, the phase transition in the interaction
of metal and Si plates. Using the calculation procedure justified
in Section 3 both thermodynamically and experimentally
(i.e., disregarding the conductivity of Si plate before and taking it
into account after the phase transition) we arrive at
\begin{equation}
{\cal F}_2(a,T)-{\cal F}_1(a,T)=-\frac{k_BT}{16\pi a^2}
\left[\zeta(3)-{\rm Li}_3(r_0)\right],
\label{eq20}
\end{equation}
\noindent
where
\begin{equation}
r_0=\frac{\varepsilon_0^{Si}-1}{\varepsilon_0^{Si}+1}\approx 0.84.
\label{eq21}
\end{equation}
\noindent
Equation (\ref{eq20}) demonstrates an abrupt change of the Casimir free
energy in the transition point. In the high-temperature limit
\begin{equation}
k_BT\gg k_BT_{\rm eff}=\frac{\hbar c}{2a}
\label{eq22}
\end{equation}
\noindent
(at room temperature of $T=300\,$K this corresponds to separations
$a>5\,\mu$m) the relative change of the free energy in accordance
with (\ref{eq20}) achieves 24\%. Thus, in the point of phase
transition the Casimir free energy and, as a consequence, the Casimir
force undergo an abrupt change although the doping concentration
and resistivity of the plate are both changing continuously.
It follows that the reflection amplitudes of real electromagnetic waves on the
plate cannot feel the phase transition that occurs with the increase
of doping concentration and,
thus, do not contain information on the respective
change of the Casimir free energy in accordance with equation (\ref{eq20}).

\section{Problems with polar dielectrics}

In the above, we have considered the dielectric permittivities in
the form of (\ref{eq4}). This form is commonly used for the
description of electronic polarization which is inherent to all
dielectrics. The respective oscillator frequencies belong to the
ultraviolet spectrum. Some dielectrics, however, contain different ions
(the typical examples are, for instance, SiO${}_2$ and Al${}_2$O${}_3$).
These dielectrics possess  {\it ionic polarization}. Their dielectric
permittivity can be also presented in the form of (\ref{eq4}) but
with oscillator frequencies belonging to the infrared spectrum. In both
cases molecules do not possess intrinsic dipole moments, but only
induced dipole moments due to the influence of the fluctuating
electromagnetic field. One more type of dielectrics is the so-called
{\it polar dielectrics} whose molecules possess intrinsic dipole moments
which are oriented in the external electromagnetic field.
In general, the dielectric permittivity of a dielectric with all
three types of polarization along the imaginary frequency axis can be
represented in the form \cite{27}
\begin{equation}
\varepsilon({\rm i}\xi)=1+\frac{f_{\rm UV}}{\omega_{\rm UV}^2+\xi^2}
+\frac{f_{\rm IR}}{\omega_{\rm IR}^2+\xi^2}+
\frac{d}{1+\xi\tau_D}.
\label{eq23}
\end{equation}
\noindent
Here, we have included for simplicity only one oscillator term
describing the electronic polarization and one oscillator term
describing the ionic polarization (the Ninham-Parsegian model).
The last term on the right-hand side of (\ref{eq23})
with the temperature dependent parameters $d$ and $\tau_D$
is the so-called
Debye term which describes the orientation polarization.
Typical values of $1/\tau_D$ belong to the microwave region of the spectrum.

\begin{figure*}[b]
\vspace*{-14.5cm}
\hspace*{-.7cm}\includegraphics{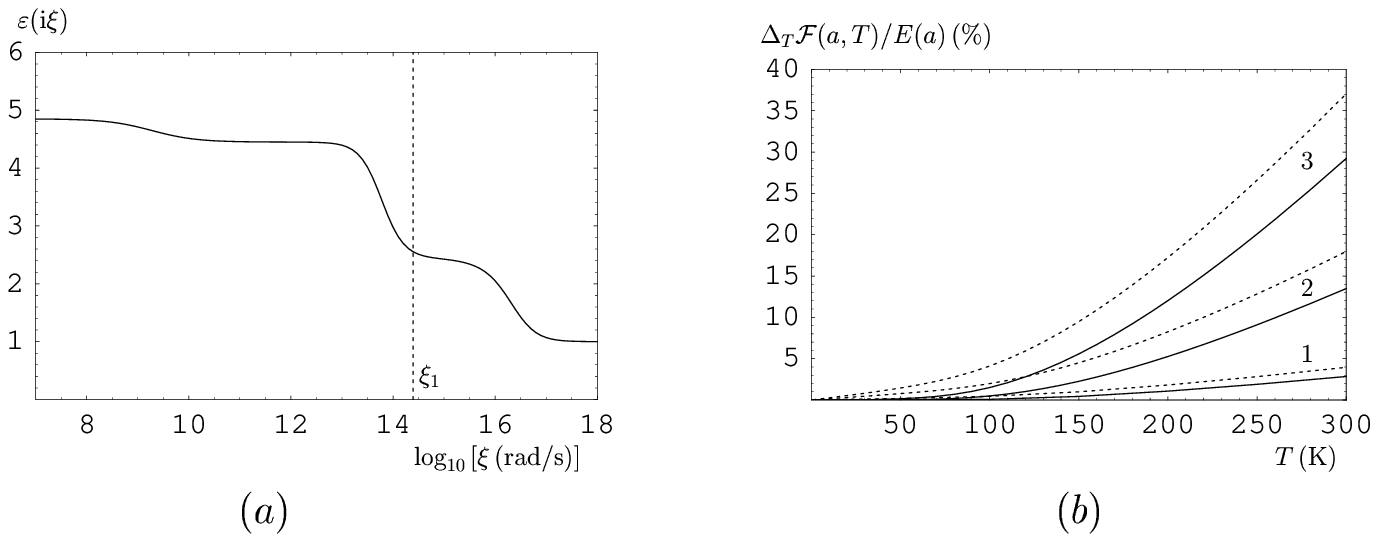}
\vspace*{-11.cm}
\caption{
(a) Dielectric permittivity of mica along the imaginary frequency axis.
The dashed line indicates the first Matsubara frequency at $T=300\,$K.
(b) The relative thermal correction to the Casimir energy of mica plates
as a function of temperature  at separations $a=100\,$nm (lines labeled 1),
$a=500\,$nm (lines labeled 2) and $a=1\,\mu$m (lines labeled 3).
Solid lines are computed by taking into account the electronic and ionic
polarization, whereas dashed lines take into account the
orientation polarization as well.
}
\end{figure*}
Let us consider mica as an example of dielectric which possesses all
the three types of polarization. The dielectric permittivity of mica along the
imaginary frequency axis is plotted in figure 4(a).
It corresponds to the following values of the parameters in (\ref{eq23}):
$\omega_{\rm UV}=10.33\,$eV, $f_{\rm UV}=157.93\,\mbox{eV}^2$,
$\omega_{\rm IR}=3.95\times10^{-2}\,$eV,
$f_{\rm UV}=3.12\times10^{-3}\,\mbox{eV}^2$ and,
at room temperature,
$\tau_D=5\times 10^{-8}\,$s, $d=0.4$ \cite{27}.
As is seen in figure 4(a), there are the three horizontal steps on the
functional dependence of $\varepsilon({\rm i}\xi)$ on
$\log_{10}\xi$ due to the three types of polarization.
The step due to the electronic polarization is in the frequency region
around $10^{15}\,$rad/s. Being extrapolated to zero frequency, this step
would lead to $\varepsilon_0^e=2.45$. The step due to
both electronic and ionic polarization
is in the frequency region of order $10^{11}-10^{12}\,$rad/s. The
extrapolation of this step to zero frequency leads to
$\varepsilon_0^{ei}=4.45$. Finally, there is the third step at frequencies
below $10^8\,$rad/s due to the electronic, ionic and orientation polarization.
As a result, the static dielectric permittivity of mica due to all three
types of polarization is equal to $\varepsilon_0^p=4.85$.

At separations below $1\,\mu$m the Casimir energy at zero temperature,
$E(a)$, is mostly determined by the electronic polarization.
It is instructive to compare the relative magnitude of the thermal
correction $\Delta_T{\cal F}(a,T)$ in (\ref{eq11}) calculated with
account of different types of polarization. Thus for Si, which
possesses the electronic polarization only,
$\Delta_T{\cal F}(a,T)/E(a)=1.45$\% at $T=300\,$K, $a=500\,$nm.
If we disregard  both ionic and orientation polarizations of mica
and take into account only its electronic polarization,
the relative thermal correction is $\Delta_T{\cal F}(a,T)/E(a)=1.25$\% at
the same $T$ and $a$. Thus, the role of the electronic polarization
of Si and mica is in fairly good agreement. However, in mica the
ionic and orientation polarizations are also present. In figure 4(b)
we plot the relative thermal correction, $\Delta_T{\cal F}(a,T)/E(a)$,
for mica versus temperature at separations $a=100\,$nm (solid and
dashed lines labeled 1),  $a=500\,$nm (lines labeled 2) and
$a=1\,\mu$m (lines labeled 3). All solid lines are computed taking the
electronic and ionic polarization into account, i.e., using the
dielectric permittivity with $\varepsilon_0^{ei}=4.45$.
The dashed lines are computed by using the complete dielectric
permittivity (\ref{eq23}), i.e., with account of the orientation polarization
also. For simplicity the room temperature values of $d$ and $\tau_D$ are
used. As is seen in figure 4(b) (solid line 2), at $T=300\,$K
the relative thermal correction achieves 13.5\% (to compare with 1.25\%
found above with account of the electronic polarization only).
Thus, the account of ionic polarization leads to a marked increase of
the relative thermal correction.

In figure 4(b) it is seen also that the role of the orientation polarization
increases with the increase of separation distance. Thus, at $a=100\,$nm,
$T=300\,$K the account of orientation polarization leads to a 1\% increase
of the relative thermal correction, but at $a=1\,\mu$m to a 8\% increase.
We emphasize that the Debye term in dielectric permittivity (\ref{eq23})
leads to problems in the Lifshitz theory. This term influences only
the zero-frequency contribution to the Casimir free energy (\ref{eq1}).
As a result, the thermal correction with account of the orientation
polarization is given by
\begin{equation}
\Delta_T{\cal F}^{(p)}(a,T)=\Delta_T{\cal F}(a,T)
-\frac{k_BT}{16\pi a^2}\left[{\rm Li}_3(r_{0,p}^2)-
{\rm Li}_3(r_{0,ei}^2)\right].
\label{eq24}
\end{equation}
\noindent
Here, $\Delta_T{\cal F}(a,T)$ is the thermal correction due to the
electronic and ionic polarizations only and
\begin{equation}
r_{0,p}=\frac{\varepsilon_0^{p}-1}{\varepsilon_0^{p}+1}, \qquad
r_{0,ei}=\frac{\varepsilon_0^{ei}-1}{\varepsilon_0^{ei}+1}.
\label{eq25}
\end{equation}
\noindent
Note that $r_{0,p}$ depends on the temperature through the
parameter $d$ in ({\ref{eq23}). At temperatures of about
$T=300\,$K (\ref{eq24}) contains a contribution being
approximately linear in the temperature.}
As to the first term on the right-hand side of (\ref{eq24}),
$\Delta_T{\cal F}(a,T)$, it has a standard form considered in \cite{15,16}.
In \cite{28} it was questioned whether or not the Debye term should be
included in the model of dielectric response used in the Lifshitz theory.
According to \cite{28}, the inclusion of the orientation degrees of
freedom that come into play at very low frequencies much below the first
Matsubara frequency is not justified. This problem calls for further
investigation.

\section{Conclusions and discussion}

In the above, we have justified the rule that for a wide range of
materials having zero conductivity at zero temperature (so-called
{\it true dielectrics}) the conductivity arising at nonzero temperature
must be disregarded in the calculation of the Casimir force using the
Lifshitz theory. These materials include not only simple insulators, but
also intrinsic semiconductors, Mott-Hubbard dielectrics and doped
semiconductors with doping concentration below critical.
We have proved that for all these materials the violation of this rule
leads to a violation of the Nernst heat theorem for the Casimir
entropy, so that the Lifshitz theory becomes thermodynamically
inconsistent. { Even more,} the inclusion of conductivity properties of
Si plate with doping concentration {below the critical one} into the model
of dielectric response was shown to be inconsistent with the data
of the recent experiment on the modulation of the Casimir force with laser
pulses \cite{10}. Thus, the proposed rule is not only warranted theoretically,
but it has already obtained  experimental confirmation.
This is a problem of great concern for the Lifshitz theory because
the inclusion of a negligible or relatively small conductivity,
arising in dielectrics at nonzero temperature, must not lead to
theoretical results significantly different of those obtained under
the neglect of this conductivity. What is more, one
could expect that the theoretical
results obtained with included conductivity are
more exact. However, in reality these results are found to be simply
invalid as being in contradiction with thermodynamics and inconsistent
with the experimental data.
Recently the modification of the Lifshitz theory of
atom-wall interaction in the high temperature limit was
suggested \cite{30} in the presence of spatial dispersion. 
The obtained interaction potential recovers the limiting
cases of dielectrics and ideal conductors with account of
low and high density of charge carriers, respectively. 
The proposed modification, however, was shown \cite{31} to violate
the Nernst theorem for a wide range of dielectrics and to be
inconsistent with measurement data of experiment \cite{10} at
a 70\% confidence level.

We have also considered the insulator-metal transition and demonstrated
that in the transition point the Casimir free energy and force undergo an
abrupt change. This may happen in an abrupt phase transition from one
crystal structure to { another one} or, alternatively, with a continuous
increase of doping concentration. In the latter case the doping
concentration and resistivity are both continuous at room temperature
in the point of phase transition. From this it follows that the reflection
amplitudes of real electromagnetic waves do not contain information
about the anomalous behavior of the Casimir force in the transition point.

One more problem {occurs when applying} the Lifshitz theory
to polar dielectrics. We have shown that the account of orientation
polarization results in large thermal correction at separations of about
hundreds of nanometers that is a linear function of the temperature.
Arguments are presented that this effect may be nonphysical.

To conclude, although there are serious problems in the application
of the Lifshitz theory to real materials, a rule can be
formulated allowing to avoid contradictions with thermodynamics and
leading to theoretical results consistent with experiment.

\section*{Acknowledgments}
The authors are greateful to U Mohideen and V M Mostepanenko
for helpful discussions.
GLK is indebted to the Center of Theoretical Studies and Institute
for Theoretical Physics, Leipzig University for kind
hospitality.
This work  was supported by Deutsche Forschungsgemeinschaft,
Grant No.~436\,RUS\,113/789/0--3.


\section*{References}
\numrefs{99}
\bibitem {1}
Casimir H B G 1948
{\it Proc. K. Ned. Akad. Wet.}
{\bf 51} 793
\bibitem{2}
Bordag M, Mohideen U and Mostepanenko V M 2001
{\it Phys. Rep.} {\bf 353} 1
\bibitem{3}
Bressi G, Carugno G, Onofrio R and Ruoso G 2002
{\it Phys. Rev. Lett.} {\bf 88} 041804
\bibitem{4}
Chen F, Mohideen U, Klimchitskaya G L and
Mos\-te\-pa\-nen\-ko V M 2002
{\it Phys. Rev. Lett.} {\bf 88} 101801 \\
Chen F, Mohideen U, Klimchitskaya G L and
Mos\-te\-pa\-nen\-ko V M 2002
{\it Phys. Rev.} A {\bf 66} 032113
\bibitem{5}
Decca R S, Fischbach E, Klimchitskaya G L,
 Krause D E, L\'opez D and Mostepanenko V M 2003
{\it Phys. Rev.} D {\bf 68} 116003
\bibitem{6}
Decca R S, L\'opez D, Fischbach E, Klimchitskaya G L,
 Krause D E and Mostepanenko V M 2005
 {\it  Ann. Phys. NY } {\bf 318} 37 \\
Klimchitskaya G L, Decca R S, L\'opez D, Fischbach E,
 Krause D E and Mostepanenko V M 2005
 {\it  Int. J. Mod. Phys.} A {\bf 28} 2205
\bibitem{7}
Decca R S, L\'opez D, Fischbach E, Klimchitskaya G L,
 Krause D E and Mostepanenko V M 2007
 {\it  Phys. Rev} D {\bf 75} 077101 \\
Decca R S, L\'opez D, Fischbach E, Klimchitskaya G L,
 Krause D E and Mostepanenko V M 2007
{\it Eur. Phys. J.} C {\bf 51} 963
\bibitem{8}
Chen F, Mohideen U, Klimchitskaya G L and
Mos\-te\-pa\-nen\-ko V M 2005
{\it Phys. Rev.} A {\bf 72} 020101(R) \\
Chen F, Mohideen U, Klimchitskaya G L and
Mos\-te\-pa\-nen\-ko V M 2006
{\it Phys. Rev.} A
{\bf 74} 022103
\bibitem{9}
Chen F,  Klimchitskaya G L,
Mos\-te\-pa\-nen\-ko V M and Mohideen U 2006
{\it Phys. Rev. Lett.}  {\bf 97} 170402
\bibitem{10}
Chen F,  Klimchitskaya G L,
Mos\-te\-pa\-nen\-ko V M and Mohideen U 2007
{\it Optics Express} {\bf 15} 4823 \\
Chen F,  Klimchitskaya G L,
Mos\-te\-pa\-nen\-ko V M and Mohideen U 2007
{\it Phys. Rev.} B {\bf 76} 035338
\bibitem{10a}
Kardar M and Golestanian R 1999
{\it Rev. Mod. Phys.} {\bf 71} 1233
\bibitem{11}
Lifshitz E M 1956
{\it Sov. Phys. JETP}  {\bf 2} 73
\bibitem{12}
Dzyaloshinskii I E, Lifshitz E M and Pitaevskii L P  1961
{\it Sov. Phys. Usp.} {\bf 4} 153 \\
Lifshitz E M and Pitaevskii L P  1984
{\it Statistical Physics}, p.II (Oxford: Pergamon)
\bibitem {13}
Bezerra V B, Klimchitskaya G L, Mostepanenko V M
and Romero C 2004
{\it Phys. Rev.} A {\bf 69} 022119
\bibitem{14}
Mostepanenko V M, Bezerra V B, Decca R S, Fischbach E, Geyer B,
Klimchitskaya G L, Krause D E, L\'opez D
and Romero C 2006
{\it J. Phys. A: Math. Gen.} {\bf 39} 6589
\bibitem{15}
 Geyer B, Klimchitskaya G L and
Mostepanenko V M
2005 {\it Phys. Rev.} D {\bf 72} 085009 \\
Klimchitskaya G L, Geyer B and
Mostepanenko V M  2006
{\it J. Phys. A: Math. Gen.} {\bf 39} 6495
\bibitem{16}
Geyer B, Klimchitskaya G L and
Mostepanenko V M
2006 {\it Int. J. Mod. Phys. } A {\bf 21} 5007 \\
Geyer B, Klimchitskaya G L and
Mostepanenko V M 2008
{\it Ann. Phys. NY} {\bf 323} 291
\bibitem{17}
Geyer B, Klimchitskaya G L and
Mostepanenko V M
2007 {\it J. Phys. A.: Mat. Theor.} {\bf 40} 13485
\bibitem{18}
Mostepanenko V M and Geyer B 2008
{\it J. Phys. A: Math. Theor.} this issue
\bibitem{20}
Palik E D (ed) 1985 {\it Handbook of Optical Constants of
Solids} (New York: Academic)
\bibitem{19}
Wilson A G 1931
{\it Proc. Roy. Soc. Lond.} A {\bf 133} 458
\bibitem{21}
Mott N F 1990
{\it Metal-Insulator Transitions} (London: Taylor and Francis)
 \bibitem{22}
Raimes S 1967
{\it The Wave Mechanics of Electrons in Metals}
(Amsterdam: North-Holland Publishing Company)
 \bibitem{23}
Shklovskii B I and Efros A L 1984
{\it Electronic Properties of Doped Semiconductors.
Solid State Series}, v.45 (Berlin: Springer)
 \bibitem{24}
Zylbersztejn A and Mott N F 1975
{\it Phys. Rev.} B {\bf 11} 4383
\bibitem{25}
Castillo-Garza R, Chang C-C, Jimenez D,  Klimchitskaya G L,
Mostepanenko V M and Mohideen U 2007
{\it Phys. Rev.} A {\bf 75} 062114
\bibitem{26}
Beadle W E, Tsai J C C and Plummer R D (eds) 1985
{\it Quick Reference Manual for Silicon Circuit Technology}
(New York: Wiley)
\bibitem {27}
Parsegian V A 2005
{\it Van der Waals forces: A Handbook for Biologists,
Chemists, Engineers, and Physicists}
(Cambridge: Cambridge University Press)
\bibitem{28}
Hough D B and White L R 1980
{\it Adv. Coll. Interface Sci.} {\bf 14} 3
\bibitem{30}
Pitaevskii L P 2008 arXiv:0801.0656
\bibitem {31}
Klimchitskaya G L, Mohideen U and Mostepanenko V M
2008 arXiv:0802.2698
\endnumrefs
\end{document}